\documentclass[preprint]{aastex62}

\usepackage{lineno}

\usepackage[]{natbib}

\usepackage{epsfig}

\usepackage{graphicx}

\usepackage{epstopdf}

\usepackage{amssymb}

\usepackage{lineno}

\usepackage{color,soul}

\usepackage{graphics}

\usepackage{graphics}

\submitjournal{ApJL}

\shorttitle{Sanchez et al. 2019}
\shortauthors{Sanchez et al. 2019}

\begin{document}


\title{Physical Characterization of Active Asteroid (6478) Gault}

\correspondingauthor{Juan A. Sanchez}
\email{jsanchez@psi.edu}

\author{Juan A. Sanchez}
\affiliation{Planetary Science Institute, 1700 East Fort Lowell Road, Tucson, AZ 85719, USA}
\altaffiliation{Visiting Astronomer at the Infrared Telescope Facility, which is operated by the University of Hawaii under Cooperative Agreement no. NNX-08AE38A with 
the National Aeronautics and Space Administration, Science Mission Directorate, Planetary Astronomy Program.}

\author{Vishnu Reddy}
\affiliation{Lunar and Planetary Laboratory, University of Arizona, 1629 E University Blvd, Tucson, AZ 85721-0092}
\altaffiliation{Visiting Astronomer at the Infrared Telescope Facility, which is operated by the University of Hawaii under Cooperative Agreement no. NNX-08AE38A with 
the National Aeronautics and Space Administration, Science Mission Directorate, Planetary Astronomy Program.}

\author{Audrey Thirouin}
\affiliation{Lowell Observatory, 1400 W Mars Hill Rd, Flagstaff, Arizona, 86001, USA}

\author{Edward L. Wright}
\affiliation{Division of Astronomy and Astrophysics, University of California Los Angeles, 430 Portola Plaza, Box 951547, Los Angeles, CA 90095-1547}

\author{Tyler R. Linder}
\affiliation{Astronomical Research Institute, 1015 Cr 1300N, Sullivan, IL 61951}
\affiliation{University of North Dakota, Clifford Hall Room 512, 4149 University Ave Stop 9008, Grand Forks, ND 58202}
\altaffiliation{Visiting Astronomer at the Infrared Telescope Facility, which is operated by the University of Hawaii under Cooperative Agreement no. NNX-08AE38A with 
the National Aeronautics and Space Administration, Science Mission Directorate, Planetary Astronomy Program.}

\author{Theodore Kareta}
\affiliation{Lunar and Planetary Laboratory, University of Arizona, 1629 E University Blvd, Tucson, AZ 85721-0092}
\altaffiliation{Visiting Astronomer at the Infrared Telescope Facility, which is operated by the University of Hawaii under Cooperative Agreement no. NNX-08AE38A with 
the National Aeronautics and Space Administration, Science Mission Directorate, Planetary Astronomy Program.}

\author{Benjamin Sharkey}
\affiliation{Lunar and Planetary Laboratory, University of Arizona, 1629 E University Blvd, Tucson, AZ 85721-0092}
\altaffiliation{Visiting Astronomer at the Infrared Telescope Facility, which is operated by the University of Hawaii under Cooperative Agreement no. NNX-08AE38A with 
the National Aeronautics and Space Administration, Science Mission Directorate, Planetary Astronomy Program.}

\begin{abstract}

Main belt asteroid (6478) Gault has been dynamically linked with two overlapping asteroid families: Phocaea, dominated by S-type asteroids, and Tamara, dominated by low-albedo C-types. This object has recently 
become an interesting case for study, after images obtained in late 2018 revealed that it was active and displaying a comet-like tail. Previous authors have proposed that the most likely scenarios to explain the observed 
activity on Gault were rotational excitation or merger of near-contact binaries. Here we use new photometric and spectroscopic data of Gault to determine its physical and compositional properties. Lightcurves derived 
from the photometric data showed little variation over three nights of observations, which prevented us from determining the rotation period of the asteroid. Using WISE observations of Gault and the near-Earth Asteroid 
Thermal Model (NEATM) we determined that this asteroid has a diameter $<$6 km. NIR spectroscopic data obtained with the Infrared Telescope Facility (IRTF) showed a spectrum similar to that of S-complex asteroids, and 
a surface composition consistent with H chondrite meteorites. These results favor a compositional affinity between Gault and asteroid (25) Phocaea, and rules out a compositional link with the Tamara family. From the spectroscopic 
data we found no evidence of fresh material that could have been exposed during the outburst episodes.

\end{abstract}



\section{Introduction} \label{sec:intro}

Asteroid (6478) Gault is a $<$10 km-sized object located in the inner part of the main belt (a$\sim$2.3 au) in the Phocaea family. This asteroid family is composed of nearly 2000 members and it is dominated by S-type 
asteroids \citep{2009MNRAS.398.1512C, 2015aste.book..297N}. Observations obtained by the Zwicky Transient Facility (ZTF) survey showed that Gault experienced two brightening events, one on 2018 October 
$18\pm5$ days and the other on 2018 December $24\pm1$ days \citep{2019ApJ...874L..16Y}. Images obtained by the Asteroid Terrestrial-impact Last Alert System (ATLAS) on 2018 December 8 revealed that Gault became 
active, displaying a 30\arcsec-long tail at a position angle PA=$290^\mathrm{o}$. On 2019 January 5, new images of Gault obtained by ATLAS showed that the tail was measuring 135\arcsec-long \citep{2019CBET}, 
and later that month a second dust tail was detected \citep{2019CBET..Jehin}. This discovery prompted \cite{2019arXiv190410530C} to look for signs of activity among NOAO archived images, which revealed that Gault 
has been active for at least six years, as seen in images taken in 2013 when a pronounced tail was already present.

Based on photometric measurements obtained by the ZTF survey, \cite{2019ApJ...874L..16Y} found that the dust ejecta was dominated by grains of up to 10 $\mu$m in size that are ejected at low velocities ($<$ 1 m s$^{-1}$), 
indicative of non-sublimation-driven ejections. They suggested that the most likely scenarios to explain the activity of this asteroid were rotational excitation or merger of near-contact binaries.

\cite{2019A&A...624L..14M}, obtained photometric data of Gault on 2019 January 13, 14, and 15 using various telescopes around the world in order to determine its rotation period. However, the lightcurves showed no 
significant variation over time. Their analysis of the dust properties yielded results consistent with those of \cite{2019ApJ...874L..16Y}.

\cite{2019ApJ...874L..20K} carried out ground-based observations of Gault from 2019 February 8 to 18. From their photometric data they determined that Gault has a rotation period of $\sim$2 h. They noticed that this value is 
close to the critical breakup limit of a rubble pile at $\sim$2.2 h per rotation \citep[e.g.,][]{2000Icar..148...12P}, suggesting that the dust emission was caused by disruption or landslides resulting from a YORP-induced rotational disturbance.

Although Gault is dynamically linked with the Phocaea family, its spectral and compositional affinity with this family have not been confirmed. Broadband colors of Gault showed that this object is more similar to C-type asteroids than 
S-types \citep{2019ApJ...874L..16Y, 2019ApJ...876L..19J}. Furthermore, \cite{2019ApJ...874L..20K} found that Gault is also dynamically linked with the low-albedo Tamara family, a recently discovered asteroid family residing in the 
Phocaea region \citep{2017AJ....153..266N}. 

In this study we present new photometric data from the IRTF and the Cerro Tololo Inter-American Observatory (CTIO) to verify the 2 h rotation period reported by \cite{2019ApJ...874L..20K}. We also constrain the diameter of Gault 
using data from the WISE mission. In addition, near-infrared (NIR) spectra of Gault obtained with the IRTF are used to determine its taxonomic type, composition, and possible meteorite analogs. We also use the spectroscopic data 
to search for signs of fresh material that could have been excavated when the asteroid became active.

\section{Physical Characterization} \label{sec:charac}

\subsection{Photometric Observations}

Gault was observed for rotational lightcurve study on three occasions over two weeks. Our first dataset was obtained through a clear filter on a 0.61 m f/6.8 telescope and an Apogee F6 camera with a pixel scale of 1.4\arcsec/
pixel and a field of view of 24$\arcmin$ square located at the CTIO facility on 2019 March 15, an exposure time of 60 s was used. On March 26, we used the MIT Optical Rapid Imaging System (MORIS) instrument on NASA IRTF 
with a pixel scale of 0.11\arcsec/pixel and a field of view of 1$\arcmin$ square. MORIS data were taken using an LPR600 filter with an exposure time of 8 s. On March 30, we collected unfiltered observations with the SMARTS 1.0 m 
telescope at CTIO using an Apogee F42 with a pixel scale of 1.05$\arcsec$ and a field of view of 9$\arcmin$ square, exposures were limited to 90 s. The CTIO datasets were both calibrated and reduced with the Canopus software 
whereas the MORIS dataset was analyzed with the techniques described in \cite{2010A&A...522A..93T}.

Two tails were seen in the images obtained at the CTIO, the first was measured to be $\sim$167\arcsec-long and the second was 5\arcsec-long. In a stack of the images taken at CTIO, we 
measured the asteroid’s Point Spread Function (PSF) away from the two tails and found it to have a full width half maximum (FWHM) of $\sim$3.05 pixels (3.2\arcsec), which is similar to the stars in the image, which have FWHM's 
of 2.95-3.2 pixels. Thus, while we see no direct evidence for an extended dust coma, if it is present it should have a lower surface brightness than either of the two tails detected.

As we did not measure any extended reflected light coming from Gault brighter than either tails, we use the brightness measured for both tails to determine their contribution as a rough estimate of the overall dust 
contamination of our spectral data. We converted a set of the CTIO images stacked on Gault to polar coordinates and measured the radial intensity profiles along both tails as well in a direction away from either tail as mentioned 
previously. We calculated the increase in brightness by subtracting the away-from-tails profile from the short and long tails. Compared to the brightest pixel of Gault, the short tail is found to be $\sim$6\% as bright and the 
long tail is found to be 1.5-2.0\% as bright in the region within several arcseconds of the asteroid. Considering the limitations of the methods, we argue that the light observed is dominated by reflected light from the asteroid’s surface 
and that the two tails make up $\sim$8\% of the light detected.

\subsection{Lightcurve Results}

\begin{figure*}[!ht]
\begin{center}
\includegraphics[height=11cm]{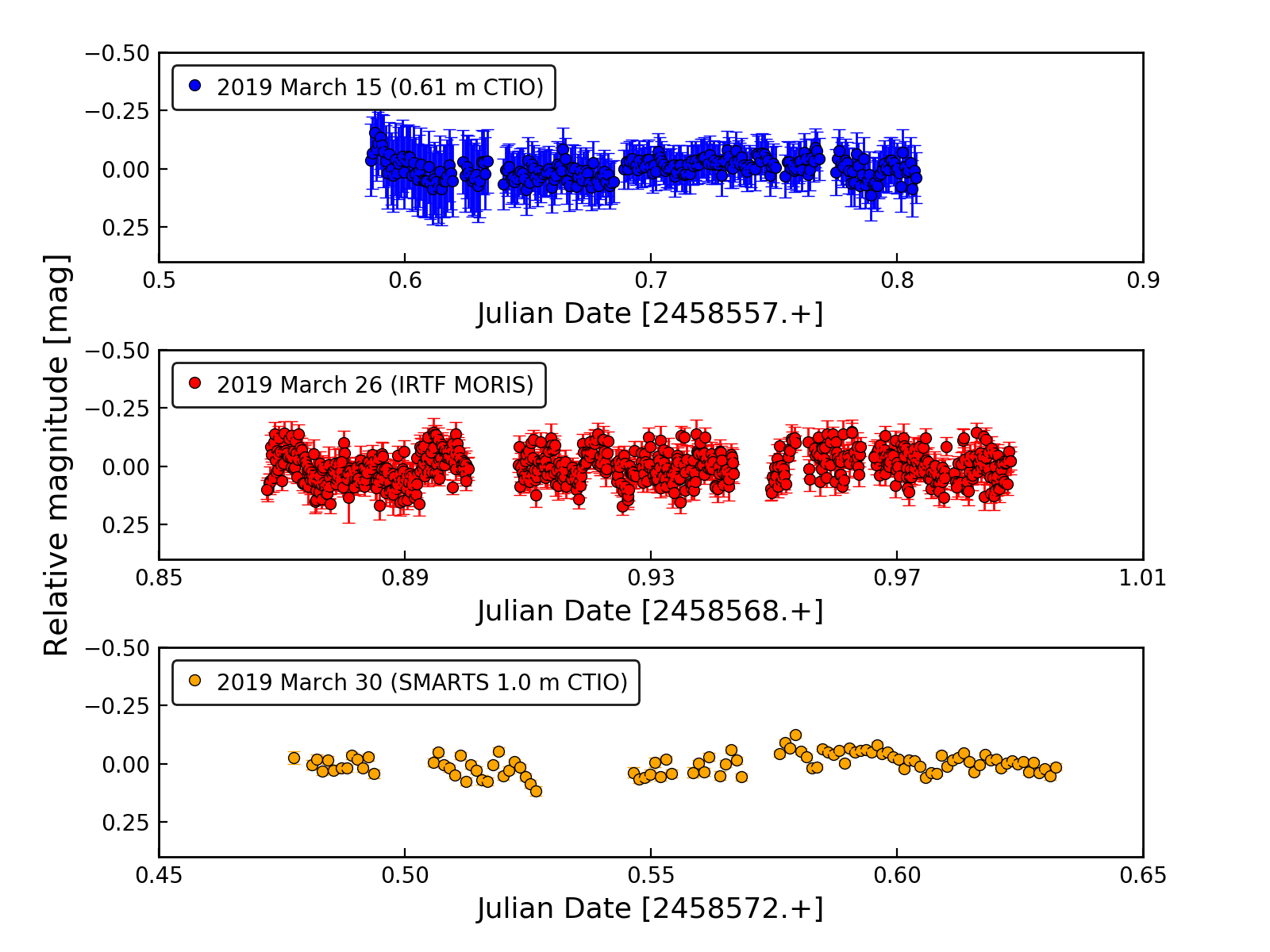}

\caption{\label{f:Figure_2} {\small Lightcurves of (6478) Gault presented in chronological order with data obtained on 2019 March 15 with the 0.61 m f/6.8 at the CTIO in blue, IRTF MORIS data obtained on March 
26 in red, and data from March 30 obtained with the SMARTS 1.0 m telescope at CTIO in orange. We find no obvious rotational period as suggested by \cite{2019ApJ...874L..20K}.}}

\end{center}
\end{figure*}

Our three datasets span $\sim$5 h, $\sim$3 h and $\sim$4 h, respectively. Cloudy conditions during the first observing run affected the quality of the photometry. All datasets show a 
flat lightcurve with no obvious variability over the duration of the observing blocks (Fig. \ref{f:Figure_2}). Our datasets were inspected for periodicities with the Lomb periodogram, and the Phase Dispersion Minimization technique 
\citep{1976Ap&SS..39..447L, 1978ApJ...224..953S}. First, we tested each dataset individually and in a second step we searched for periodicity using the whole sample. In all cases, no rotational period was favored with 
a high enough ($>99.9\%$) confidence level.

As stated earlier, the lightcurve of Gault was also studied in \cite{2019A&A...624L..14M} and \cite{2019ApJ...874L..20K}. Our flat lightcurve is consistent with the \cite{2019A&A...624L..14M} results. Based on our data, we cannot 
confirm the rotational period of $\sim$2 h for Gault (assuming a double-peaked lightcurve) reported by \cite{2019ApJ...874L..20K}.  These results are compatible with an asteroid observed pole-on, or an object having a 
spherical shape. As discussed by \cite{2019A&A...624L..14M}, the presence of dust around the asteroid would further complicate the analysis of a lightcurve with a small amplitude.

\subsection{Thermal Modeling}

The only WISE observations of Gault in the Minor Planet Center database are 15 detections centered on 2016 August 25.6 when it was near perihelion and only 1.39 au from the Earth.  But a search of the WISE image
archive shows 21 frames covering the position of Gault.  Clearly the flux was close to the detection limit, so positive noise fluctuations would get detected while negative fluctuations would not.  To avoid bias, a PSF fit to the pixels 
surrounding Gaul on all 21 frames was done, yielding W1 = 15.912$\pm$0.091 mag and W2 = 13.856$\pm$0.033 mag. The heliocentric distance was $r = 1.89$ au and the phase angle was $\alpha = 31.8^\circ$.  A similar search 
of the cryogenic phases of the WISE mission showed 16 frames during the 3-band cryo phase centered on 2010 August 26.9.  PSF fits to the pixels in these frames give W1=17.72$\pm$0.32 mag, W2=16.11$\pm$0.33 mag, and 
W3=9.00$\pm$0.24 mag. Gault was $\Delta = 2.38$ au from Earth, $r = 2.57$ au from the Sun, with a phase angle $\alpha = 23.1^\circ$ during these observations.

A thermal model NEATM \citep{1998Icar..131..291H} fit to these WISE data plus an absolute magnitude ($H$) of 14.4$\pm$0.3 was performed, but the combination of the low SNR in the W2-W3 color and the uncertain 
contribution of reflected 
light to the W2 flux required the addition of prior knowledge to stabilize the fit.  We ran a Bayesian calculation using a Monte Carlo Markov chain having log normal priors with $p_{IR}/p_V =  1.76\pm34\%$ and $\eta = 1.2\pm29\%$.  
The posterior of the fit had a geometric albedo $p_V = 0.176\pm44\%$, $p_{IR}/p_V =  1.39\pm24\%$, $\eta = 1.36\pm18\%$, and a diameter $D = 3.96 \pm 22\%$ km.  A caveat is that the diameter and beaming are tightly 
correlated in the posterior with $D \propto \eta^{1.2}$, and the prior contributes 39\% of the weight in the $\eta$ value, so the $\eta$ prior still has an important effect on the diameter value. The distribution of $\eta$ seen in 
\cite{2014ApJ...791..121M} has a median of 0.9 and a 1$\sigma$ range of $\pm13\%$ with only 5 out of 3080 objects showing $\eta$ greater than the +1$\sigma$ point ($\eta = 1.615$) in the posterior for Gault.  Thus, a diameter
greater than the posterior +2$\sigma$ point of 6.1 km is unlikely.

\subsection{Spectroscopic Observations}

We carried out observations of Gault on 2019 March 26 UTC with the SpeX instrument \citep{2003PASP..115..362R} on the IRTF. NIR spectra (0.7-2.5 $\mu$m) of the asteroid, extinction and solar analog stars were 
obtained in low-resolution (R$\sim$150) prism mode with a 0.8$\arcsec$ slit width. All spectra were obtained at the parallactic angle to minimize differential refraction at the shorter wavelength end.  

A total of thirty 200 s spectra of Gault were acquired when the asteroid was 17.8 visual magnitude, at a phase angle of 14$^\mathrm{o}$. A G-type local extinction star was observed before and after the asteroid 
observations in order to correct the telluric absorption bands. NIR spectra of solar analog star SAO 120107 were also obtained to correct for spectral slope variations that could be introduced by the use of a non-solar local extinction star. Data reduction was performed using the IDL-based software Spextool \citep{2004PASP..116..362C}. A detailed description of the data reduction procedure is presented in \cite{2013Icar..225..131S, 2015ApJ...808...93S}.

\subsection{Compositional Analysis}

\begin{figure*}[!ht]
\begin{center}
\includegraphics[height=11cm]{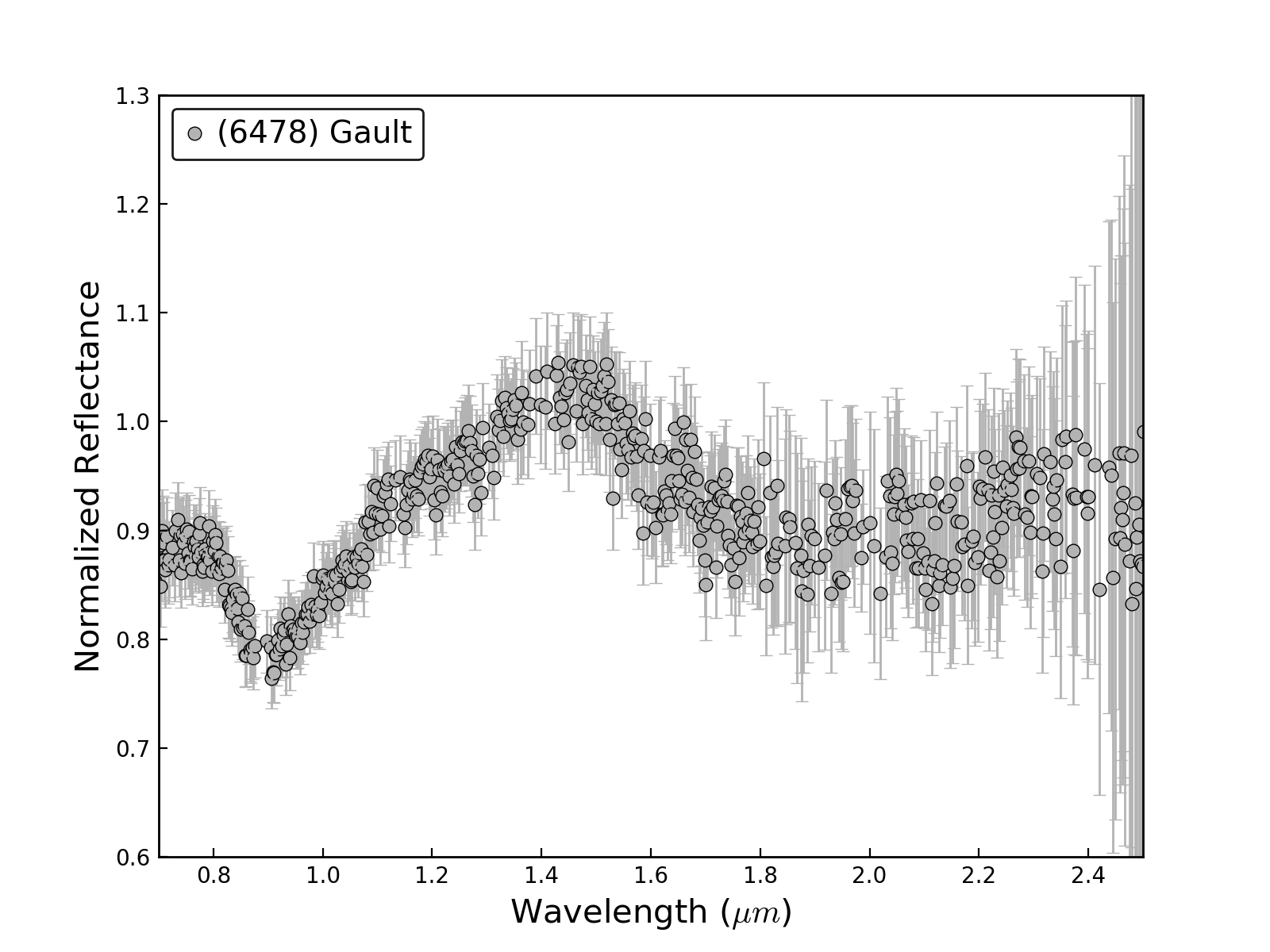}

\caption{\label{f:Figure_3} {\small Near-IR spectrum of (6478) Gault obtained using the SpeX instrument on NASA IRTF. The spectrum exhibits two absorption bands at $\sim$ 1- and 2-$\mu$m due to the presence of olivine and pyroxene. The spectrum is normalized to unity at 1.5 $\mu$m.}}

\end{center}
\end{figure*}

The NIR spectrum of Gault is shown in Fig. \ref{f:Figure_3}.  Two absorption bands at $\sim$ 1- and 2-$\mu$m due to the presence of olivine and pyroxene can be seen. We have performed the taxonomic classification of the 
asteroid using the online Bus-DeMeo taxonomy calculator (http://smass.mit.edu/busdemeoclass.html), which showed that Gault belongs to the S-complex, specifically to the Sr-type (PC1' = 
0.1636, PC2' = 0.0679) under this taxonomic system \citep{2009Icar..202..160D}. Sr-types typically exhibit a deeper 2-$\mu$m band than S-types, however we noticed that the large scattering of the data in this 
band is likely affecting the PC2' value, which increases as this band becomes deeper, thus it is also possible that the object is an S-type. 

\begin{figure*}[!ht]
\begin{center}
\includegraphics[height=11cm]{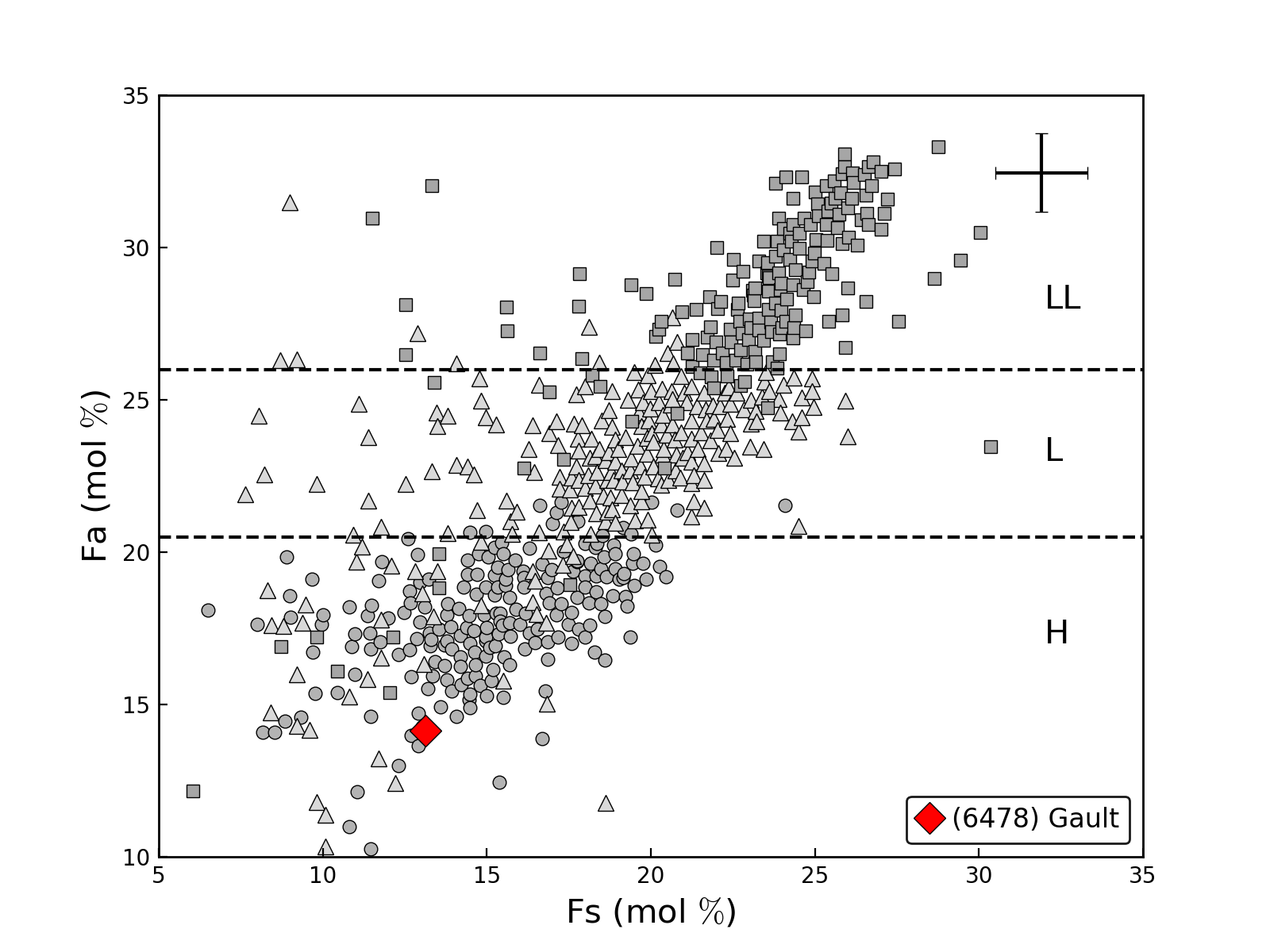}

\caption{\label{f:Figure_4} {\small Iron abundance in silicate minerals on Gault represented as mol\% of fayalite (Fa) vs. ferrosilite (Fs). Measured values for LL (squares), L (triangles), and H (circles) ordinary chondrites 
from \cite{2011ScienceNakamura} are also included. The error bars in the upper right corner correspond to the uncertainties derived by \cite{2010Icar..208..789D}, 1.3 mol\% for Fa, and 1.4 mol\% for Fs. Figure adapted 
from \cite{2011ScienceNakamura}.}}

\end{center}
\end{figure*}

Spectral band parameters including band centers, band depths and Band Area Ratio (BAR), as well as their associated errors were measured using a Python code following the procedure 
described in \cite{2012Icar..220...36S}. We used the measured Band I center (0.92$\pm$0.01 $\mu$m) and the equations of \cite{2010Icar..208..789D} to determine the olivine and pyroxene chemistry, which are given by 
the molar contents of fayalite (Fa) and ferrosilite (Fs), respectively. The BAR can be used to calculate the olivine-pyroxene abundance ratio (ol/(ol+px)), however, the scattering of the data in the 2-$\mu$m band could make this 
parameter unreliable, therefore we decided not to use it. The olivine and pyroxene chemistries of Gault were found to be Fa$_{14.1\pm1.3}$ and Fs$_{13.1\pm 1.4}$, respectively. Figure  \ref{f:Figure_4} shows the 
molar content of Fa vs. Fs for Gault and measured values for LL, L, and H ordinary chondrites from \cite{2011ScienceNakamura}. Within the uncertainties, these values fall in the range of those found for H chondrites by 
\cite{2010Icar..208..789D}. This result rules out a compositional affinity between Gault and the Tamara family, whose members are thought to be low-albedo ($<$ 0.1) C-types, and would favor a link between Gault and 
asteroid (25) Phocaea, whose composition has been found to be consistent with ordinary chondrites \citep{2019AALN}. Since S-type asteroids are composed of anhydrous silicates, a volatile-driven activity for Gault 
seems unlikely.

\subsection{Looking for fresh excavated material}

\begin{figure*}[!ht]
\begin{center}
\includegraphics[height=11cm]{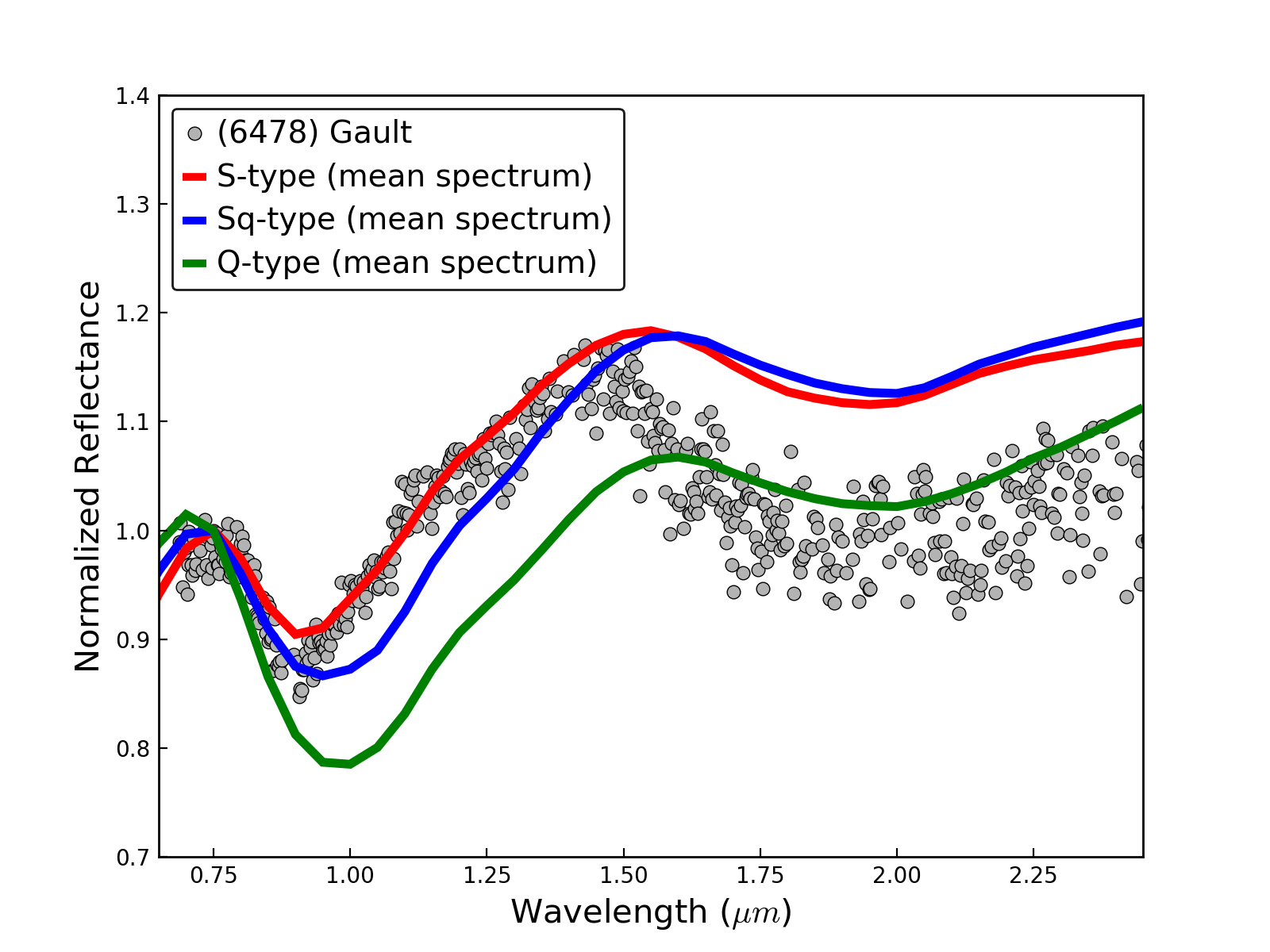}

\caption{\label{f:Figure_5B} {\small Comparison between the NIR spectra of (6478) Gault and the mean spectra of Q-, Sq-, and S-types from \cite{2009Icar..202..160D}. All spectra are normalized to unity at $\sim$ 0.75 
$\mu$m.}}

\end{center}
\end{figure*}

Different mechanisms have been proposed to explain dust ejection from asteroids, including volatile sublimation, rotational mass loss, impacts, and thermal disintegration among others \citep{2015aste.book..221J}. As stated 
earlier, in the case of Gault, merger of near-contact binaries or rotational breakup are thought to be the physical mechanisms behind the activity of the asteroid \citep{2019ApJ...874L..16Y, 2019ApJ...874L..20K}. Regardless of what 
led to the activation of Gault, it would be reasonable to think that landslides and dust ejection could have excavated fresh material from beneath the surface of the asteroid.
The NIR spectrum of this fresh material would exhibit deeper absorption bands and a relatively flat spectral slope compared to that of the weathered material. Ideally, the best way to look for signs of fresh exposed 
material would be to compare spectra obtained before and after the asteroid became active, unfortunately, no pre-outburst spectral data of Gault exist. However, this limitation can be overcome by comparing the spectrum of Gault with those 
of Q-, Sq- and S-type asteroids, as these taxonomic types are thought to represent a weathering gradient, with Q-types having relatively fresh surfaces, and Sq- and S-types having more space-weathered surfaces 
\citep[e.g.,][]{2010Natur.463..331B, 2019Icar..324...41B, 2018AJ....155..140R}. Figure \ref{f:Figure_5B} shows the spectrum of Gault and the mean spectrum of a Q-, Sq-, and a S-type asteroid from \cite{2009Icar..202..160D}. We 
found that the Band I depth of Gault (14.3$\pm$0.5\%) is much closer to that of the S-type (13.0$\pm$0.2\%) than the Sq- (17.5$\pm$0.3\%) and Q-type (23.8$\pm$0.1\%). We have also used the 
principal components PC1' and PC2' obtained in the previous section to calculate the Space Weathering Parameter $\Delta\eta$ \citep{2010Natur.463..331B}. This parameter is given by the scalar magnitude of the space 
weathering vector defined in the principal component space of the Bus-DeMeo taxonomy \citep{2010Natur.463..331B}, and can be used to estimate the degree of space weathering experienced by an asteroid. For Gault, we 
determined a value $\Delta\eta=0.608$, consistent with an extensively weathered surface. These results suggest three possible scenarios: 1) no fresh material was excavated; 2) fresh material was excavated but we cannot detect 
it; 3) our current models for asteroid space weathering are incomplete. In the first scenario, the surface of Gault would have reached a state of saturated space weathering. According to \cite{2019Icar..324...41B}, under this scenario, 
surface grains can become uniformly weathered after multiple re-arrangements events followed by extended periods of exposure to the space environment, hence causing saturation by space weathering. Thus, when a new 
re-arrangement of the regolith occurs during a resurfacing event, no fresh material will be exposed. In the second scenario, it is possible that fresh material was excavated in localized regions too small to be detected. 
\cite{2019A&A...624L..14M} estimated that the total mass ejected during the 2018 and 2019 events was equivalent to a spherical volume of $\sim$10 m radius. Since the NIR spectrum obtained with the IRTF is a 
disk-integrated spectrum, the contribution of such a small region to the overall spectral features would be negligible. In addition, we cannot rule out the possibility that the exposed areas were covered by re-accumulation of a 
mixture of fresh and weathered debris. This mechanism has been proposed to explain the lack of spectral variations seen in the site of fission of primary asteroids in asteroid pairs \citep{2014Icar..243..222P}.  Within the 
second scenario we also have to consider the possibility that dust could be temporarily masking the spectral signature of the fresh material. However, as discussed in section 2.1, no evidence for an extended coma was found 
and it is not clear whether the contribution from the tails would be enough for this to happen. Finally, regarding the third scenario, it is also possible that the tools that we are employing to quantify space weathering on 
asteroids cannot be generalized to all objects, even if they belong to the same taxonomic type. This idea seems to be supported from spacecraft observations of (433) Eros and (243) Ida (both S-types) that show different space 
weathering trends on these asteroids \citep{2010Icar..209..564G}.

\section{Summary} \label{sec:summ}

We have obtained photometric and NIR spectroscopic data of active asteroid (6478) Gault in an effort to independently confirm previous results and to determine for the first time the composition of this asteroid. Our results 
can be summarized as follows: 

\begin{itemize}

\item From the lightcurves we were unable to confirm the $\sim$2 h rotation period determined by \cite{2019ApJ...874L..20K}. Our results are consistent with those of \cite{2019A&A...624L..14M}. 

\item A thermal model fit to WISE data yielded values of $p_V = 0.176\pm44\%$, and $D=3.96 \pm 22\%$ km for the geometric albedo and diameter, respectively.

\item The olivine and pyroxene chemistries of Gault were found to be consistent with those of H-type ordinary chondrites, suggesting a compositional affinity with (25) Phocaea.

\item We found no sign of fresh material that could have been exposed during the outburst episodes. This can be seen in the overall shape of the spectrum and the intensity of the 1-$\mu$m band of Gault, which are more 
similar to those of an S-type than Sq- and Q-types.

\end{itemize}

\acknowledgments

This research work was supported by NASA Near-Earth Object Observations Grant NNX17AJ19G (PI: Reddy). We thank the IRTF TAC for awarding time to this project, and to the IRTF TOs and MKSS staff for their 
support. The authors wish to recognize and acknowledge the very significant cultural role and reverence that the summit of Mauna Kea has always had within the indigenous Hawaiian community. We are most fortunate 
to have the opportunity to conduct observations from this mountain. AT is partially supported by Lowell Observatory funds. Taxonomic type results presented in this work were determined, in whole or in part, using a Bus-DeMeo 
Taxonomy Classification Web tool by Stephen M. Slivan, developed at MIT with the support of National Science Foundation Grant 0506716 and NASA Grant NAG5-12355. We thank the anonymous referee for the thorough review 
and comments that helped improve the manuscript.



\end{document}